\documentclass[pre,onecolumn]{revtex4}
\usepackage{amssymb}
\usepackage[dvips]{graphicx}

\begin{document}

\preprint{}
\title[]{Identification of arches in 2D granular packings}
\author{Roberto Ar\'{e}valo}
\affiliation{Departamento de F\'{\i}sica, Facultad de Ciencias, Universidad de Navarra,
E-31080 Pamplona, Spain.}
\author{Luis A. Pugnaloni}
\affiliation{Instituto de F\'{\i}sica de L\'{\i}quidos y Sistemas Biol\'{o}gicos
(UNLP-CONICET), cc. 565, 1900 La Plata, Argentina.}
\author{Diego Maza}
\affiliation{Departamento de F\'{\i}sica, Facultad de Ciencias, Universidad de Navarra,
E-31080 Pamplona, Spain.}

\begin{abstract}
We identify arches in a bed of granular disks generated by a molecular
dynamic-type simulation. We use the history of the deposition of the
particles to identify the supporting contacts of each particle. Then, arches
are defined as sets of mutually stable disks. Different packings generated
through tapping are analyzed. The possibility of identifying arches from the
static structure of a deposited bed, without any information on the history
of the deposition, is discussed.
\end{abstract}

\volumeyear{}
\volumenumber{}
\issuenumber{}
\eid{}
\date{7 March 2006}
\startpage{1}
\endpage{}
\maketitle

\section{Introduction}

Arches are multi-particle structures encountered in non-sequentially
deposited granular beds. In some sense, arches are to granular packings what
clusters are to colloids: they represent collective structures whose
existence and behavior determine many properties of the system as a whole.
During the settlement of a granular bed, some particles come to rest
simultaneously by contacting and supporting each other. These mutually
stabilized sets of particles are called arches or bridges \cite{Mehta1}. Of
course, these arches are themselves supported by some of the surrounding
particles in the system: the arch bases. Typically, 70\% of the particles
are part of arches in a granular packing of hard spheres \cite{Pugnaloni1};
and around 60\% in packings of hard disks \cite{Pugnaloni3}. Arch formation
is crucial in the jamming of granular flows driven by gravity \cite%
{To1,To2,Zuriguel1,Zuriguel2} and it has also been proposed as a mechanism
for size segregation \cite{Durand1,Durand2}.

Arching is the collective process associated to the appearance of voids that
lower the packing fraction of the sample. Moreover, arching is directly
related to the reduction of particle--particle contacts in the assembly,
which determines the coordination number \cite{Pugnaloni1,Pugnaloni2}. In
2D, for example, the mean support number $\langle z\rangle _{\text{support}}$%
---i.e., the coordination number that accounts only for the contacts that
serve as support of at least one grain---can be obtained from the arch size
distribution $n(s)$ as \cite{Pugnaloni3}

\begin{equation}
\langle z\rangle _{\text{support}}=2\left[ 1+\frac{1}{N}\sum_{s=1}^{N}n(s)%
\right] .  \label{eq1}
\end{equation}%
Here $N$ is the total number of particles in the packing and $n(s)$ is the
number of arches consisting of $s$ grains. The number of particles that do
not form part of any arch corresponds to $n(s=1)$. This mean support
number coincides with the mean coordination number $\left\langle
z\right\rangle $ when the packing does not have any non-supporting contacts.
In non-sequentially deposited beds---especially for soft particles---there
will be contacts that are not essential to the stability of any of the two
touching particles.

A systematic study of arching is particularly complex despite the simplicity
associated to the concept of arch. The typical structural properties of
granular materials are mainly connected with the topological complexity of
the contact network. Concepts like \textquotedblleft force
chain\textquotedblright\ and \textquotedblleft force
propagation\textquotedblright\ have been profusely studied \cite{books}, but
many open questions still remain. The statistical properties of the arching
process could give some insight into these complex problems.

Arches form dynamically during non-sequential particle deposition. Given a
settled granular sample, identifying if two particles belong to the same
arch is often impossible without knowing the history of the deposition
process. Assuming that we are able to identify (experimentally) the contacts
of each particle, it is impossible to know which of these contacts are
responsible for supporting each particle in place against gravity. For
convex hard particles in 2D, only two of the contacts of a given particle
provide its stability (in 3D this number is three). The supporting contacts
are the first two (three in 3D) contacts made by the given particle that
provide stability against the external force that drives deposition (e.g.
gravity). Any contact made afterwards will not provide the essential
stability assuming that the already formed supporting contacts persist. Of
course, if one removes a supporting contact, it is possible that a
non-supporting contact of the particle becomes a supporting contact, hence
some non-supporting contacts may provide secondary (alternative) stability
to the particle.

Finding the supporting contacts of a particle is simple in a pseudo dynamics
simulation approach since these contacts are the essence of the simulation
algorithm \cite{Pugnaloni1,Pugnaloni2,Pugnaloni3,Manna1}. However, in a
realistic granular simulation ---as well as in experiments--- one needs to
track the particle collisions and analyze stability at every instant to
catch the supporting contacts of every particle in the moment they first
appear.

In this work, we report a numerical study on the supporting contacts that
are created during the non-sequential deposition of a granular bed. Then,
arches are straightforwardly obtained and analyzed. We use molecular
dynamics simulations and restrict ourselves to a 2D system of inelastic
particles in order to reduce the computational cost. We show that our
identification of supporting contacts is useful in deciding when the packing
has settled, which is a major issue in these type of simulations. We compare
our results with those obtained by pseudo dynamics methods. We also show
that using simple criteria it is possible to identify arches to some degree
without knowing the deposition history. This is of particular interest for
experimental studies where tracking particles at high velocities is rather
expensive.

\section{The 2D soft-particle molecular dynamics approach}

We simulate the process of filling a box with grains by using a
soft-particle two-dimensional molecular dynamics (MD). In this case, the
particles (monosized disk) are subjected to the action of gravity.
Particle--particle interactions are controlled by the particle--particle
overlap $\xi =d-\left\vert \mathbf{r}_{ij}\right\vert $ and the velocities $%
\dot{\mathbf{r}}_{ij}$, $\omega _{i}$ and $\omega _{j}$. Here, $\mathbf{r}%
_{ij}$ represents the center-to-center vector between particles $i$ and $j$, 
$d$ is the particle diameter and $\omega $ is the particle angular velocity.
These forces are introduced in the Newton's translational and rotational
equations of motion and then numerically integrated by standard methods \cite%
{Schafer1}. 

The contact interactions involve a normal force $F_{\text{n}}$ and a
tangential force $F_{\text{t}}$. In order to simplify the calculus we use a
normal force which involves a linear (Hookean) interaction between particles.

\begin{equation}
F_{\text{n}}=k_{\text{n}}\xi -\gamma _{\text{n}}v_{i,j}^{\text{n}}
\label{normal}
\end{equation}

\begin{equation}
F_{\text{t}}=-\min \left( \mu |F_{\text{n}}|,|F_{\text{s}}|\right) \cdot 
\text{sign}\left( \zeta \right)  \label{tangent}
\end{equation}%
where

\begin{equation}
F_{\text{s}}=-k_{\text{s}}\zeta -\gamma _{\text{s}}v_{i,j}^{\text{t}}
\label{contact}
\end{equation}

\begin{equation}
\zeta \left( t\right) =\int_{t_{0}}^{t}v_{i,j}^{\text{t}}\left( t^{\prime
}\right) dt^{\prime }  \label{static}
\end{equation}

\begin{equation}
v_{i,j}^{\text{t}}=\dot{\mathbf{r}}_{ij}\cdot \mathbf{s}+\frac{1}{2}d\left(
\omega _{i}+\omega _{j}\right)  \label{vtan}
\end{equation}

The first term in Eq.~(\ref{normal}) corresponds to a restoring force
proportional to the superposition $\xi $ of the interacting disks and the
stiffness constant $k_{n}$. The second term accounts for the dissipation of
energy during the contact and is proportional to the normal component $%
v_{i,j}^{\text{n}}$ of the relative velocity $\dot{\mathbf{r}}_{ij}$ of the
disks. The restitution coefficient is an exponentially decaying function of
the dissipation coefficient $\gamma _{\text{n}}$ \cite{Schafer1}, i.e., an
increase in the dissipation coefficient leads to a nonlinear decline of the
restitution coefficient. 

Equation~(\ref{tangent}) provides the magnitude of the force in the
tangential direction. It implements the Coulomb's criterion with an
effective friction following a rule that selects between static or dynamic
friction. Dynamic friction is accounted for by the friction coefficient $\mu 
$. The static friction force $F_{\text{s}}$ [see Eq. (\ref{contact})] has an
elastic term proportional to the relative shear displacement $\zeta $ and a
dissipative term proportional to the tangential component $v_{i,j}^{\text{t}%
} $ of the relative velocity. In Eq. (\ref{vtan}), $\mathbf{s}$ is a unit
vector normal to $\mathbf{r}_{ij}$. The elastic and dissipative
contributions are characterized by $k_{\text{s}}$ and $\gamma _{\text{s}}$
respectively. The shear displacement $\zeta $ is calculated through Eq. (\ref%
{static}) by integrating $v_{i,j}^{\text{t}}$ from the beginning of the
contact (i.e., $t=t_{0}$). The tangential interaction behaves like a damped
spring which is formed whenever two grains come into contact and is removed
when the contact finishes \cite{wolf}.

The model presented above includes all features that turned out to be
essential in the work here reported. If only dynamic friction forces are
used in the tangential direction, these keep changing slightly \emph{but
continuously} once the disks have been deposited, making it impossible to
decide whether the deposition process has definitively finished. The
addition of static friction forces allows the deposit to reach a stable
configuration. Furthermore, apart from the energy dissipation on normal
contact, the tangential component is also responsible for energy dissipation
through the second term in Eq.~(\ref{contact}). This leads to a fast decay
of the total kinetic energy of the system which final value is negligible in
comparison with simulations without tangential dissipation.

\section{Arch identification}

To identify arches one needs first to identify the two supporting particles
of each disk in the packing. Then, arches can be identified in the usual way 
\cite{Pugnaloni1,Pugnaloni2,Pugnaloni3}: we first find all \emph{mutually
stable particles} ---which we define as directly connected--- and then we
find the arches as chains of connected particles. Two disks $i$ and $j$ are 
\emph{mutually stable} if $i$ supports $j$ and $j$ supports $i$.

Regardless of the way grains are deposited, they meet with and split up from
other grains until they come to rest. While this process takes place, each
time a new contact happens we can check if any of the two touching particles
may serve to the stability of the other and save this information. Equally,
we can update this data when touching grains come apart. Eventually, when
all particles have settled, we should find that every particle has two
supporting contacts and that no updates take place any longer. The algorithm
that we follow to detect and update supporting contacts is described below.

We consider that two particles in contact with particle $i$ may provide
support to it only if the two particles are one at each side of $i$ and the
center of mass ($cm$) of particle $i$ is above the segment that joins the
two contacts. We call \emph{contact chord} to this segment. For a particle
and a wall to support particle $i$ we just need to take into account that
the contact between the wall and $i$ has the same $y$-coordinate as $i$. A
particle in contact with the base is stable per se.

If at any time a particle $i$ has a single contact, we consider this contact
as a potential first stabilizing contact (awaiting for the second
stabilizing contact to occur) only if $i$ has a higher $y$-coordinate than
the contacting particle. A contacting wall cannot be considered a first
stabilizing contact.

We construct two arrays [$nR(i)$ and $nL(i)$ with $i=0,...N-1$] that store
the indices of the \emph{right} and \emph{left} supporting particle of all
particles in the system. If a particle has an \emph{undefined support} the
corresponding position in the array is set to $-2$. If one of the supporting
contacts is a \emph{container's wall} (or base) the corresponding position
in the array is set to $-1$. Initially, all elements of $nR$ and $nL$ are
set to $-2$. After an update of all particle positions in the simulation, we
check the status of $nR$ and $nL$, and update them according to the protocol
described below. The protocol is repeated until no changes are induced in $%
nR $ and $nL$ before a new simulation time step is advanced.

\begin{figure}[tbp]
\includegraphics{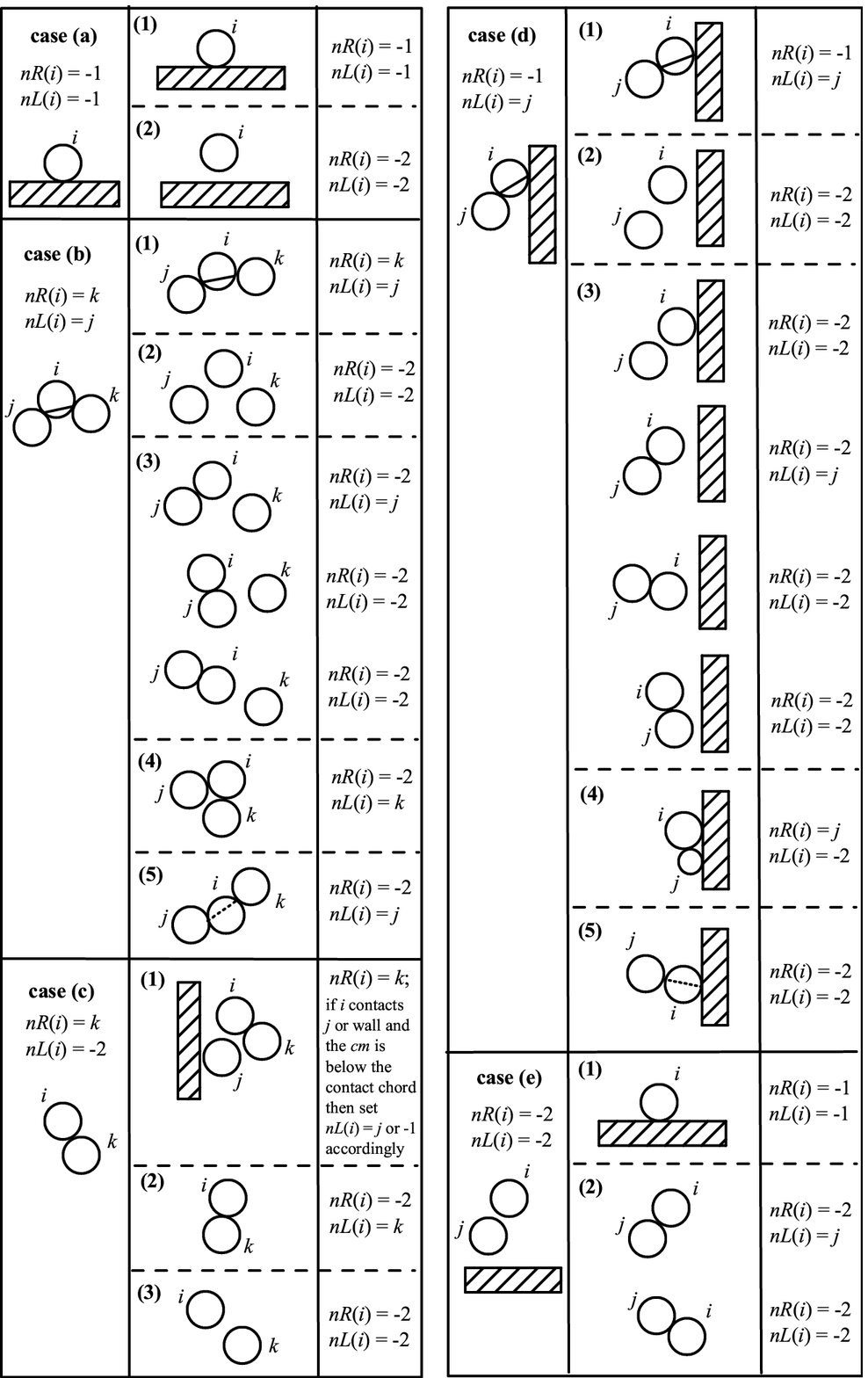}% Here is how to import EPS art
\caption{Schematic representation of the possible states of stability that a
particle $i$ can have at any given time (left column) and update protocol
that is followed according to the particle position after a simulation step
(right columns). The arrays $nR(i)$ and $nL(i)$ store the indices of the
disks that support disk $i$ (see text). In case (d-4) we draw a small disk $j
$ since this situation is difficult to appreciate by eye for equal-sized
disks where small differences in disk-to-wall overlaps are responsible for
particle $j$ moving to the right of $i$.}
\label{fig:fig1}
\end{figure}

There are six different situations that may arise depending on the stability
status of a particle previous to the position update. Within each of these
cases there are different actions to take depending on the new positions of
the particles after the position update. A summary of the most relevant
situations is shown in Fig.~\ref{fig:fig1}. Not all plausible situations are
considered here because some do not occur in practice or are extremely rare.

\emph{Case (a):} Here, particle $i$ was resting on the container's base
before the position update. We simply check if the particle is still in
contact with the base; if not, $nR(i)$ and $nL(i)$ are set to $-2$.

\emph{Case (b): }In this case, particle $i$ was supported (before the
position update took place) by other two particles which indices are stored
in $nR(i)$ and $nL(i)$. We have to check if these particles are still
supporting $i$; if not, the supporting particles must be redefined according
to these possible situations:

(1) $nR(i)$ and $nL(i)$ are still in contact with $i$, $nR(i)$ is still to
the right of $i$ and $nL(i)$ is still to the left of $i$, and the $cm$ of $i$
is above the contact chord $\rightarrow $ leave $nR(i)$ and $nL(i)$
unchanged.

(2) neither $nR(i)$ nor $nL(i)$ are in contact with $i$ $\rightarrow $ set $%
nR(i)$ and $nL(i)$ to $-2$.

(3) either $nR(i)$ or $nL(i)$ is no longer in contact with $i$ $\rightarrow $
set the lost contact to $-2$ and check that the remaining contacting
particle is still on its side (right or left) and below particle $i$; if
not, set also this contact to $-2$.

(4) $nR(i)$ and $nL(i)$ are still in contact with $i$, but either $nR(i)$ is
not to the right of $i$ or $nL(i)$ is not to the left of $i$ $\rightarrow $
say $nR(i)$ is not to the right of $i$; then (i) set $nL(i)$ to the
contacting particle [$nR(i)$ or $nL(i)$] with lower $y$-coordinate, (ii) set 
$nR(i)=-2$.

(5) $nR(i)$ and $nL(i)$ are still in contact with $i$, $nR(i)$ is still to
the right of $i$ and $nL(i)$ is still to the left of $i$, but the $cm$ of $i$
is not above the contact chord $\rightarrow $ set $nR(i)$ or $nL(i)$ to $-2$%
, the one with highest $y$-coordinate.

\emph{Case (c):} This case corresponds to a particle with a first
potentially stabilizing contact. Let us assume that this corresponds to $%
nR(i)$. We have to check that $nR(i)$ is still in place and if any new
contact can complete the stability condition. The following situations may
arise:

(1) $nR(i)$ is still in contact with $i$, $nR(i)$ is still to the right of $%
i $ $\rightarrow $ look for any particle $j$ (or wall) contacting $i$ on its
left side such that the $cm$ of $i$ is above the contact chord; if any is
found, then set $nL(i)=j$.

(2) $nR(i)$ is still in contact with $i$, but $nR(i)$ is not to the right of 
$i$ $\rightarrow $ set $nL(i)=nR(i)$ and $nR(i)=-2$.

(3) $nR(i)$ is not in contact with $i$ $\rightarrow $ set $nR(i)=-2$.

\emph{Case (d):} This case is similar to case (b) but one of the two
container's walls takes the place of one of the supporting particles. Let us
assume that $nR(i)=-1$, i.e. we are considering the right wall. Again, we
have to redefined supporting contacts according to these possible situations:

(1) $nR(i)$ and $nL(i)$ are still in contact with $i$, $nL(i)$ is still to
the left of $i$, and the $cm$ of $i$ is above the contact chord $\rightarrow 
$ leave $nR(i)$ and $nL(i)$ unchanged.

(2) neither $nR(i)$ nor $nL(i)$ are in contact with $i$ $\rightarrow $ set $%
nR(i)$ and $nL(i)$ to $-2$.

(3) either $nR(i)$ or $nL(i)$ is no longer in contact with $i$ $\rightarrow $
(i) set the lost contact to $-2$; (ii) if the remaining contact is $nR(i)$
(the wall), then set $nR(i)=$ $-2$, else check that $nL(i)$ is still to the
left of $i$ and with a $y$-coordinate below particle $i$; if not, set also
this contact to $-2$.

(4) $nR(i)$ and $nL(i)$ are still in contact with $i$, but $nL(i)$ is not to
the left of $i$ $\rightarrow $ set $nR(i)=nL(i)$ and set $nL(i)=-2$.

(5) $nR(i)$ and $nL(i)$ are still in contact with $i$, $nL(i)$ is still to
the left of $i$, but the $cm$ of $i$ is not above the contact chord $%
\rightarrow $ set $nR(i)$ and $nL(i)$ to $-2$.

\emph{Case (e):} This case corresponds to a particle $i$ that was in the air
(without defined supporting contacts) before the position update. We need to
check if any contact has occurred and if it is a potential first stabilizing
contact. Before doing this, we check that the particle has negative vertical
velocity to avoid considering particles that are in their way up after a
bounce. Two situations may arise:

(1) Particle $i$ contacts the \ container's base $\rightarrow $ set $nR(i)$
and $nL(i)$ to $-1$.

(2) Particle $i$ contacts another particle $j$ $\rightarrow $ if $i$ has
higher $y$-coordinate than $j$, then set either $nR(i)$ or $nL(i)$
(depending on which side of $i$ is $j$) to $j$ .

\emph{Case (f):} This case never arises because a contacting wall cannot be
considered a first stabilizing contact.

\section{Results on tapped granular beds}

In order to test the algorithm for the identification of supporting contacts
and arches in a realistic non-sequential deposition we carry out MD
simulations of the tapping of $512$ disks in a rectangular box $13.91d$
wide. The values used for the parameters of the force model are: $\mu =0.5$, $k_{\text{n}}=10^{5}$, $\gamma _{\text{n}}=300$, $k_{%
\text{s}}=\frac{2}{7}k_{\text{n}}$ and $\gamma _{\text{s}}=200$ with an
integration time step $\delta =10^{-4}$. The stiffness constants $k$ are measured in units of $mg/d$, the
damping constants $\gamma $ in $m\sqrt{g/d}$ and time in $\sqrt{d/g}$. Here, 
$m$, $d$ and $g$ stand, respectively, for the mass of the disks, the
diameter of the disks and the acceleration of gravity. The walls and base are represented by
disks with infinite mass and diameter. The velocity-Verlet method was used
to integrate the equations of motion along with a neighbor list to speed up
the simulation. In order to keep stable the numerical integration, a basic requirement is to use an integration time step $\delta $ much smaller than the contact time between particles, which is essentially
controlled by $\gamma _{\text{n}}$. We have chosen $\delta $ to be 50 times shorter than the typical duration of a contact. A more detailed discussion about the numerical algorithm can be found in Ref. \cite{ludingnato}. 

The value of $\gamma _{%
\text{n}}$ is chosen deliberately high in order to have a small coefficient
of normal restitution ($e_{\text{n}}=0.058$). This way we reduce computer
time since fast energy dissipation leads the system to rest in less
time. Of course, this also reduces the number of bounces and oscillations in the system.

Disks are initially placed at random without overlaps in the simulation box, and the initial velocities are assigned from a Gaussian distribution of mean zero. The $y$-coordinates range from $0$ up to $100d$. Then, particles are allowed to deposit under the action of gravity. The deposit is said to be stable when the supporting contacts arrays $nR(i)$ and $nL(i)$ (see Sec. III), remain unchanged for $10^{4}$ time-steps. Then, the stable configuration is saved. It is worth
mentioning that this criterion to decide when the system is ``at rest" is
very reliable. All our simulation runs end up with unchanging supporting
contacts arrays. Even though particles may perform small vibrations about
their equilibrium positions and orientations, they remain supported by the
same set of particles and this is what defines the mechanical stability in a
macroscopic sense.

\begin{figure}[tbp]
\includegraphics{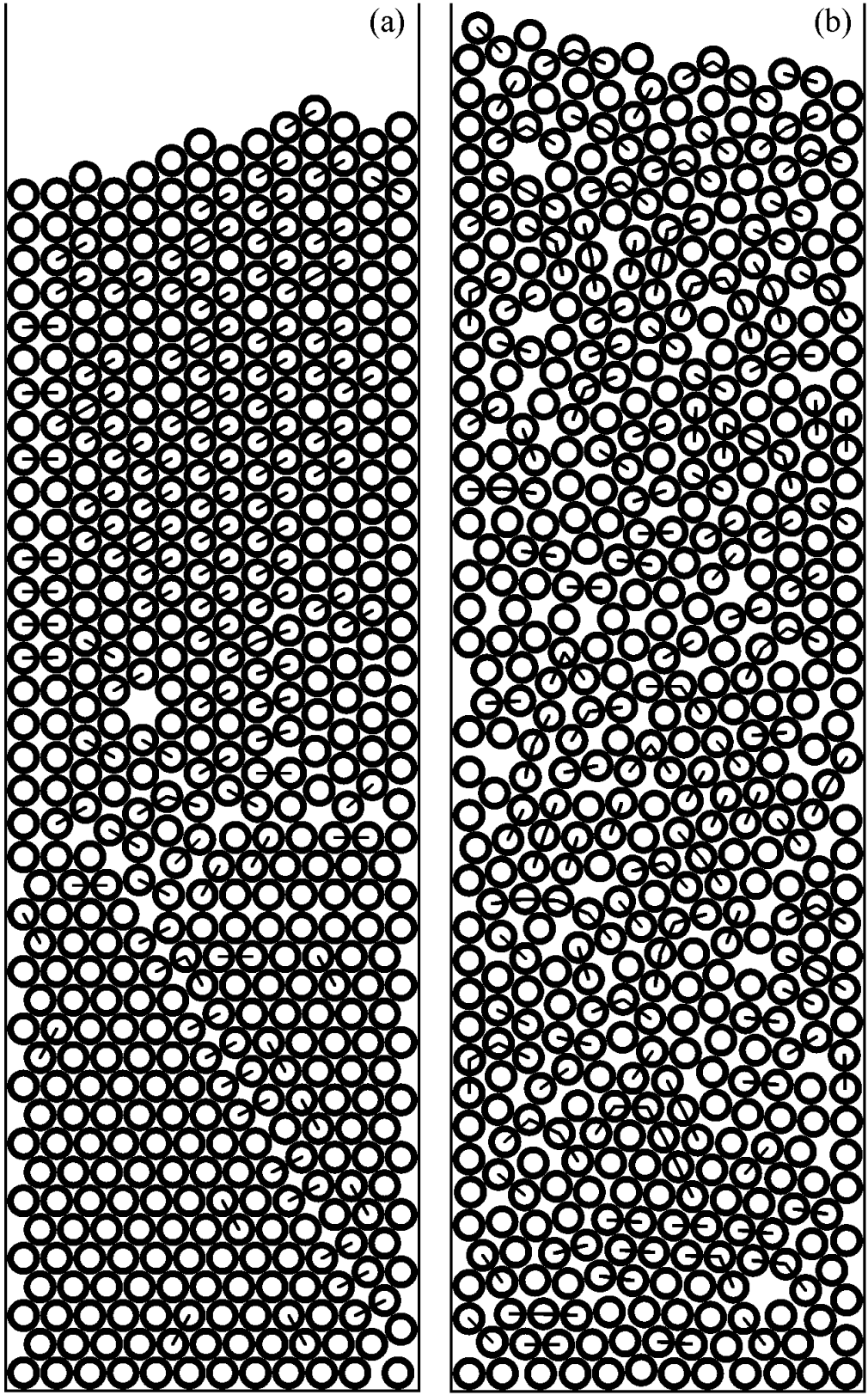}% Here is how to import EPS art
\caption{Two snapshots of the configurations obtained for tapping amplitudes 
$\Gamma =0.71$ (a) and $4.99$ (b). Arches identified with the protocol of
Sec. III are indicated by segments that join disks belonging to the same
arch. Note that some disks may seem to be in unstable positions since all
contacts are above its center. These disks are in fact stable thanks to the
static friction forces. For each arch there are two disks that
form the base of the arch, these are not indicated in the figure.}
\label{fig:fig2}
\end{figure}

The tapping process is simulated by moving the container's base and walls in
the vertical direction half period of a sine function of amplitude $A$ and
frequency $w=2\sqrt{g/d}$. The tapping amplitude is measured through the peak acceleration $\Gamma =Aw^{2}$. After all particles have settled down
and the stable configuration has been saved, we increase the amplitude of
the perturbation by $\Delta \Gamma =0.0134g$ and the process is repeated.
The range of the tapping amplitude goes from $0$ to $6.41g$. When the maximum amplitude is reached the system is then tapped with decreasing amplitudes down to $\Gamma =0$. Then, a new increasing $\Gamma$ cycle is started and so on.

The ``annealing" protocol is performed several times starting from different
initially deposited beds and then the results are averaged separating the
very first increasing $\Gamma$ phase from the rest of the tapping ramps. We introduce this averaging method due to the fact that all quantities obtained for
the initial increasing ramp do not match the subsequent decreasing and increasing ramps for amplitudes below some threshold $\Gamma _{0}$ [See inset of Fig. 3(b)]. The first increasing ramp of the
tapping protocol is known as the \emph{irreversible branch } and the
subsequent coinciding ramps are known as the \emph{reversible} branch. Despite the
introduction of static friction (which could induce history dependent
effects) and the fact that we apply a single tap to the system at each $%
\Gamma $, the results show the same trends as many experiments \cite%
{Nowak1,rennes,makse} where the granular assembly is subjected to some sort
of tapped annealing. Unfortunately, the slow decaying time of the compaction
process makes virtually impossible to simulate a real experimental situation.

In order to enhance the quality of the averaged arch properties at given
states along the tapping ramp we take intermediate configurations ($\Gamma
=0.71$ and $4.99$) and tap the system at constant $\Gamma $ for longer runs (%
$1000$ cycles). These values of $\Gamma $ correspond to a high density and a
low density state of the system. The results obtained from these longer runs
are essentially the same as those obtained from the corresponding states
along the annealing process; only the statistical dispersion is improved.

We can see two snapshots of configurations obtained at low and high tapping
amplitudes in Fig.~\ref{fig:fig2}. The arches identified in the sample can
be appreciated by means of the segments that join mutually stabilizing disks
(see Sec. III). A single disconnected segment indicates that the two joined
disks support each other and therefore form an arch. A chain of connected
segments indicates that all joined disks belong to the same arch. For each
arch there are two disks that form the base of the arch, these are not
indicated in the figure. The configuration with $\Gamma =0.71$ [Fig.~\ref%
{fig:fig2}(a)] is quite ordered, showing a localized disordered region at
middle height in the packing. Large crystal-like domains of triangular order
are observed with clear defect boundaries in agreement with experiments \cite%
{Rankenburg1}. Here, most arches consist of only two particles; only a few
three-particle arches can be appreciated. Figure~\ref{fig:fig2}(b)
corresponds to $\Gamma =4.99$. This configuration is clearly more disordered
than the former; however, void spaces are rather uniform in size and
homogeneously distributed throughout the packing. This contrasts with the
larger degree of disorder displayed by pseudo dynamic simulations \cite%
{Pugnaloni3} where a wider distribution of void sizes is found. Figure~\ref%
{fig:fig2}(b) shows a larger number of arches than Fig.~\ref{fig:fig2}(a),
particularly larger arches consisting of up to six disks.

\begin{figure}[tbp]
\includegraphics{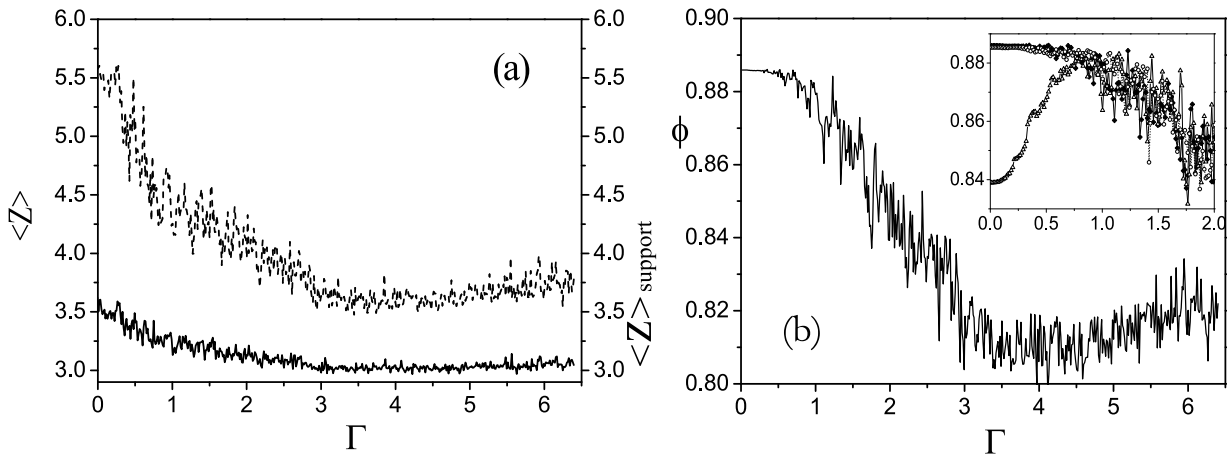}% Here is how to import EPS art
\caption{Coordination number, support number and packing fraction as a
function of the tapping amplitude $\Gamma $ within the reversible branch of the annealing protocol. (a) Coordination number $%
\left\langle z\right\rangle $ (dashed line) and support number $\left\langle
z\right\rangle _{\text{support}}$ (solid line). (b) Packing fraction $%
\protect\phi $. The inset in panel (b) shows separately the curves corresponding to several tapping ramps including the irreversible branch (lower curve).}
\label{fig:fig3}
\end{figure}

In Fig.~\ref{fig:fig3} we show the results of the mean coordination number $%
\left\langle z\right\rangle $ and packing fraction $\phi $ obtained for the
reversible branch of the tapping ramp (i.e., decreasing ramp). We also show
the values for $\left\langle z\right\rangle _{\text{support}}$ that includes
only those contacts that serve to the stability of at least one of the two
touching particles. Unlike hard particles \cite{Pugnaloni3}, soft particles present more
contacts than those just needed to make particles stable, therefore $%
\left\langle z\right\rangle _{\text{support}}$ is not related to $%
\left\langle z\right\rangle $ in a trivial way. We have to bear in mind that
is $\left\langle z\right\rangle _{\text{support}}$ which is directly related
to the arch size distribution and not $\left\langle z\right\rangle $. Our
results show that $\left\langle z\right\rangle _{\text{support}}$ and $%
\left\langle z\right\rangle $ present qualitatively the same behavior
although $\left\langle z\right\rangle _{\text{support}}$ presents lower
values and varies within a narrower range.

We can see from Fig.~\ref{fig:fig3} that both $\left\langle z\right\rangle $
and $\phi $ have rather high values when the tapping amplitude is small,
corresponding to an ordered system. As $\Gamma $ is increased those values
undergo a slow decrease as the system gets disordered. Finally, $\phi $ and $%
\left\langle z\right\rangle $ present a slight increase from their minimum
values for high tapping amplitudes. This result qualitatively agrees with
the experimentally observed behavior in 3D \cite{Nowak1} where a minimum in
the density dependence on $\Gamma $ has been observed. However, simulations
of disks through pseudo dynamics \cite{Pugnaloni3} show a much sharper
transition between the ordered (low $\Gamma $) and the disordered (high $%
\Gamma $) regime. Moreover, these simulations \cite{Pugnaloni3} present a
positive slope in $\left\langle z\right\rangle _{\text{support}}$ for low
tapping amplitudes within the ordered regime.

\begin{figure}[tbp]
\includegraphics{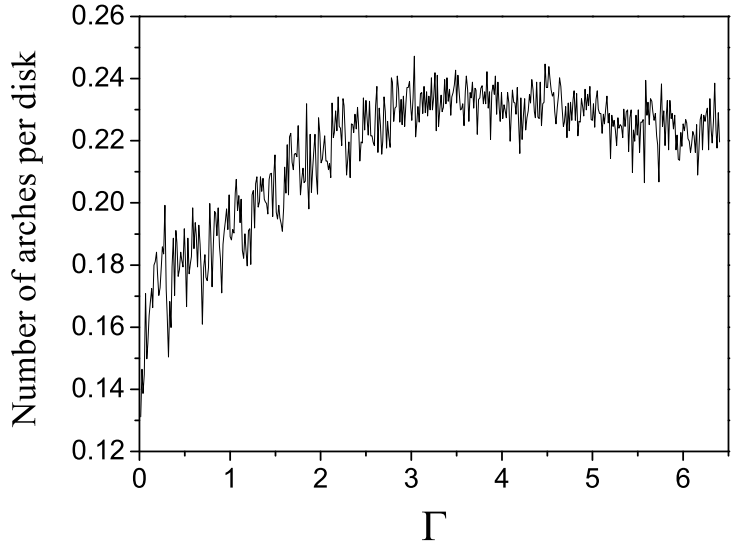}% Here is how to import EPS art
\caption{Number of arches per disk as a function of $\Gamma $. Only the
reversible (decreasing) branch of the tapping ramp is shown.}
\label{fig:fig4}
\end{figure}

Let us look into the relationship between the former quantities and the
arches identified by our algorithm. Fig.~\ref{fig:fig4} shows the number of
arches normalized by the number of particles as a function of $\Gamma $
along the reversible branch. We see that the large decrease in $\left\langle
z\right\rangle $ and $\phi $ (Fig.~\ref{fig:fig3}) corresponds to an
increase of the total number of arches in the bed. This number reaches a
maximum value and then shows a slight decrease in correspondence with the
increase in $\left\langle z\right\rangle $ and $\phi $ for high values of $%
\Gamma $. Clearly, there is a correlation between the buildup(breakdown) of
arches and the number of contacts and the free volume encountered in the
sample. At very low tapping amplitudes the particles deposit in a very
ordered manner without forming many arches; this means that, according to
Eq. (\ref{eq1}), $\left\langle z\right\rangle _{\text{support}}$ is rather
high ($\approx 3.5$, being $4$ the maximum allowed value in 2D); moreover,
there are few arch-trapped voids and $\phi $ is high. At moderate $\Gamma $,
arches are created with high probability; the coordination number then falls
accordingly and the arch-trapped voids lower the packing fraction. For very
high tapping amplitudes arches are again broken down partially; new
particle--particle contacts are created and arch-trapped voids are filled up.

\begin{figure}[tbp]
\includegraphics{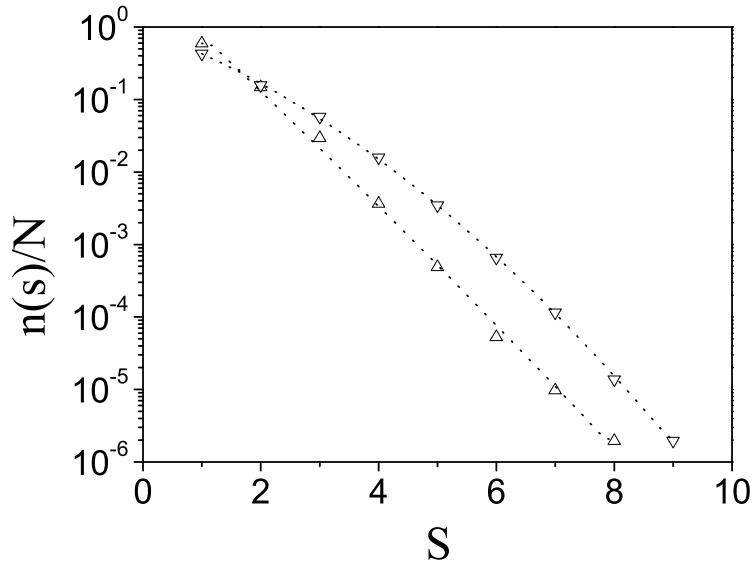}% Here is how to import EPS art
\caption{Distribution of arch sizes for $\Gamma =4.99$ (down triangles) and $%
0.71$ (up triangles). The lines are fits to a second order polynomial. The
data for $s=1$ correspond to the number of disks that are not forming arches.
}
\label{fig:fig5}
\end{figure}

In Fig.~\ref{fig:fig5}, the arch size distribution $n(s)$ is shown for two
values of the tapping amplitude. We can observe that for moderate tapping
amplitudes the system has a larger number of large arches whereas for gentle
tapping there is a larger amount of disks that do not belong to any arch ($%
s=1$). Interestingly, in both cases the distribution can be fitted to a
second order polynomial, in a semilogarithmic scale. This is in agreement
with results from pseudo dynamics simulations \cite{Pugnaloni3}. In 3D it
has been found that $n(s)\varpropto s^{-1.0\pm 0.03}$ \cite%
{Pugnaloni1,Pugnaloni2} which corresponds to a significantly higher
prevalence of large arches with respect to 2D packings.

\begin{figure}[tbp]
\includegraphics{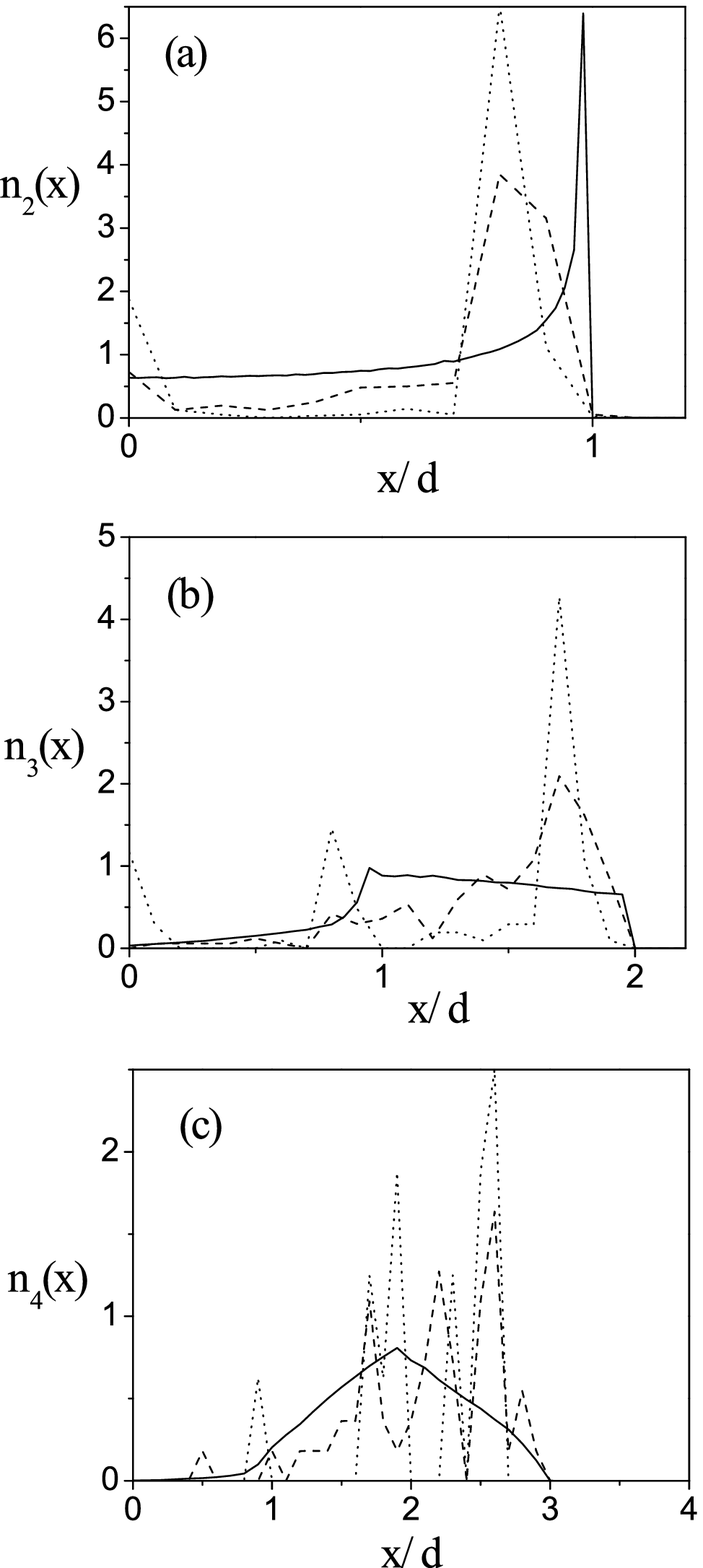}% Here is how to import EPS art
\caption{Horizontal span distribution for arches with $2$ (a), $3$ (b) and $%
4 $ (c) disks. The lines correspond to $\Gamma =0.71$ (dotted) and $\Gamma
=4.99$ (dashed). The solid lines correspond to the restricted random walk
model \protect\cite{To1}}
\label{fig:fig6}
\end{figure}

In Fig.~\ref{fig:fig6} we show the distributions of the horizontal span of
the arches and compare them with a theoretical model \cite{To1} based on a
restricted random walk. The horizontal span is the projection onto the
horizontal axis of the segment that joins the centers of the right-end disk
and the left-end disk of an arch. We use the results obtained for arches
composed by two, three and four disks obtained for two different tapping
amplitudes. At high packing fractions arch extensions appear discretized.
Since the system presents an ordered layered structure, the particles of any
arch of $s$ disks form a string that connects layers (or stay in the same
layer) so that the end-to-end horizontal projection of the arch can only be
an integer number of on-layer jumps plus an integer number of between-layers
jumps; both jumps have characteristic horizontal projections in the
triangular lattice formed by the disks. We saw above that although $\phi $
undergoes a decrease in its value for intermediate tapping amplitudes, it
remains rather high. For this reason, although horizontal span distributions
are more homogeneous, we still find some characteristic peaks for high $%
\Gamma $. These results do not agree with the prediction of the random walk
model where horizontal span distributions are smooth. However, it must be
taken into account that the restricted random walk model was proposed to
represent arches at the outlet of a hopper where particles cannot order but
simply fit in the gap between walls.

\section{Arches from static configurations}

In this section we intend to show to what extent an attempt to identify
arches from the static structure of a deposited system can yield realistic
results. We test two criteria that select two of the contacting particles of
each particle as the supporting pair of the given grain: (a) \emph{random
stabilizing pair}, and (b) \emph{lowest stabilizing pair}. We identify all
pairs of contacting particles that may---according to their relative
positions and the contact chord criterion---stabilize a given grain. For
criterion (a) we then choose one of these pairs at random as the supporting
pair. For criterion (b) we choose the pair that has the lower center of
mass. In principle, we expect that case (b) should provide a more reliable
identification of supporting contacts since those contacting particles in
lower positions may correspond to the ``real" supports, whereas those in
higher positions may correspond to contacts made by further deposition of
other particles on top of the particle under consideration. We will see that
this is not the case.

\begin{figure}[tbp]
\includegraphics{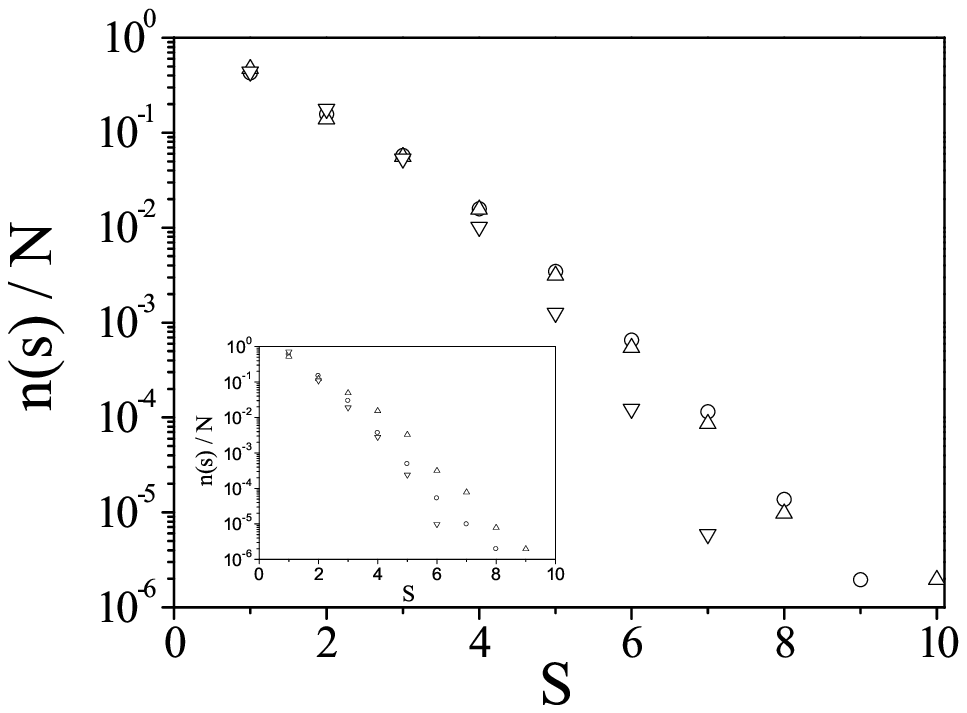}% Here is how to import EPS art
\caption{Distribution of arch sizes for $\Gamma =4.99$ comparing the ``real"
distribution obtained from simulations (circles) with the distributions
obtained from the static structure using criteria (a) (up triangles) and (b)
(down triangles). The inset shows the same results for $\Gamma =0.71$.}
\label{fig:fig7}
\end{figure}

We have obtained the arch size distribution $n(s)$ by applying the static
criteria above to packings corresponding to $\Gamma =0.71$ and $4.99$. Fig.~%
\ref{fig:fig7} shows clearly that, for disordered packings ($\Gamma =4.99$),
arches are identified much better by the random criterion. The lowest
supporting pair criterion detects less arches than the realistic dynamic
criterion. As we can see \ from the inset of Fig.~\ref{fig:fig7}, for very
ordered packings ($\Gamma =0.71$), both criteria fail. Criterion (a)
overestimates whereas criterion (b) underestimates the number of arches
beyond $s=2$. A detailed analysis of the supporting contacts detected by the
static criteria shows that criterion (a) tends to find many false mutually
stabilizing pair of particles in ordered packings, which leads to the
identification of false arches. On the other hand, criterion (b) tends to
detect too few of the real mutually stabilizing pairs of particles and for
that reason the number of arches detected are fewer than with the dynamic
criterion.

\section{Conclusions and final remarks}

We have presented a protocol to dynamically identify arches during the
deposition of granular particles carried out through realistic granular
dynamics. We find that simple static criteria can correctly identify arches
to some degree for not very ordered packings. In particular, the criterion
we call \emph{random stabilizing pair} seems to be the most successful and
could be used to identify arches in experimentally generated 2D granular
beds \cite{Blumenfeld1,Rankenburg1}. However, in ordered packings is very
difficult to identify arches from static configurations. The reason for this
is that in an ordered packing the coordination number is rather high and
then the number of plausible stabilizing pairs for a given particle is
higher than that in disordered packings. Since choosing the correct pair is
the essence of the criterion to successfully identify stabilizing contacts,
the more pairs we have to chose from, the more likely we make a mistake.

The packings generated by varying the tapping amplitude ``quasistatically"
present a low-density disordered regime (at high tapping amplitudes) and a
high-density, ordered regime (at low tapping amplitudes). The support number
decreases with increasing tapping amplitude except for of the very high
tapping amplitudes. These observations are in contrast with pseudo dynamic
simulations \cite{Pugnaloni3}\ where $\langle z\rangle _{\text{support}}$
increases with $\Gamma $ (except at the order--disorder transition). We
believe that this is due to the fact that the pseudo dynamics is a good
representation for fully inelastic disks that roll without slip and that
``fall down" at constant velocity---a situation that seems to be met by
particles carried on a conveyor belt at low velocities \cite{Blumenfeld1}.
The simulations in the present work are rather far from this regime since
particles can spring away from an impact and so explore a wider range of
position in the box before finding a locally stable configuration.

The form of the arch size distributions is in agreement with pseudo dynamic
simulations. This suggests that the form of $n(s)$ is rather insensitive to
the deposition algorithm. The calculated number of arches grows sharply as
the density decreases when the tapping amplitude is increased. However, this
tendency is not monotonic and above a transition zone the number of arches
decreases whereas the density increases. This feature highlights the
intuitive connection between arches and density fluctuations.

We have to point out here that the criterion we used to decide if two
contacts may be the stabilizing contacts for a given particle $i$ (i.e. that
the contact chord is below the $cm$ of $i$), rules out the possibility that
a particle be supported from ``above". This situation does indeed arises
occasionally due to static friction. Two contacting particles with $y$%
-coordinates slightly above particle $i$ may sustain the particle due to
high static friction forces. We have found that this situation indeed occurs
in our simulations. Indeed, some instances can be appreciated in Fig.~\ref%
{fig:fig2}(b). A detailed analysis of these types of arches that are also
found in experiments (see for example Ref. \cite{To1}) will be presented
elsewhere.

\begin{acknowledgments}
The authors thanks Gary C. Barker and Angel Garcimart\'{\i}n for useful
discussions. This work has been supported by CONICET of Argentina and also
partially by the project FI2005-03881 MEC (Spain), and PIUNA, University of
Navarra. RA thanks Friends of the University of Navarra for a grant.
\end{acknowledgments}

\end{document}